\documentclass{PoS}

\usepackage{amsmath}
\usepackage{amsfonts}
\usepackage{cite}
\usepackage{mciteplus}

\newcommand{\dd}[1]{\mathrm{d}{#1}}
\newcommand{\order}[1]{\mathcal{O}\left(#1\right)}
\newcommand{\braket}[2]{\langle #1 \mid #2 \rangle}
\newcommand{\ket}[1]{\mid #1 \rangle}

\newcommand{\Sig}[1]{\hat{\sigma}^\text{#1}}
\newcommand{\siG}[1]{\hat{\sigma}_\text{#1}}
\newcommand{\sig}[2]{\hat{\sigma}^\text{#1}_\text{#2}}

\title{Sector-improved residue subtraction: Improvements and Applications\thanks{Preprint number: TTK-18-33}}

\ShortTitle{Sector-improved residue subtraction}

\author{\speaker{A. Behring}, M. Czakon, R. Poncelet\\
        Institut f{\"u}r theoretische Teilchenphysik und Kosmologie,
        RWTH Aachen University,\\ D-52056 Aachen, Germany\\
        E-mail: \email{behring@physik.rwth-aachen.de}}

\abstract{We discuss two recent developments of the sector-improved residue
          subtraction scheme for handling real radiation at NNLO in QCD. We
          present a new phase space construction which minimizes the number
          phase space configurations for subtraction terms and we rederive
          the four-dimensional formulation of the scheme.}

\FullConference{Loops and Legs in Quantum Field Theory (LL2018)\\
		29 April 2018 - 04 May 2018\\
		St. Goar, Germany}

\begin{document}

\section{Introduction}

The increasing precision delivered by the experiments at the LHC puts the
Standard Model to more and more stringent tests. To keep up with the shrinking
experimental uncertainties, theory predictions for Standard Model processes also
need to be calculated at higher orders in perturbation theory. For many
processes this means that they have to be calculated at next-to-next-to-leading
order (NNLO) of quantum chromodynamics (QCD). While a growing number of
processes is already available at this order -- and some at even higher orders
-- a level of automation comparable to the situation at next-to-leading order
(NLO) has not yet been reached.

One of the ingredients necessary for NNLO QCD predictions is a procedure to deal
with infrared singularities of additional real emissions, which cancel against
corresponding singularities from virtual corrections for infrared safe
observables \cite{Kinoshita:1962ur,Lee:1964is}. Since the phase space integrals
are usually solved via numerical integration, these infrared divergences have to
be isolated and cancelled before the numerical treatment. Over the last few
decades, a number of schemes and techniques have been developed for this task.
At NLO the most commonly used general schemes are the Catani-Seymour dipole
subtraction \cite{Catani:1996vz,Catani:1996jh}, the Frixione-Kunszt-Signer (FKS)
scheme \cite{Frixione:1995ms,Frederix:2009yq} and the Nagy-Soper scheme
\cite{Nagy:2003qn,Bevilacqua:2013iha,Czakon:2015cla}.
Beyond NLO, there has been a lot of activity in the development and application
of general NNLO schemes, which includes
Antenna Subtraction
\cite{GehrmannDeRidder:2005cm,Currie:2013vh,GehrmannDeRidder:2007jk,%
      GehrmannDeRidder:2004tv,GehrmannDeRidder:2005hi,GehrmannDeRidder:2005aw,%
      Gehrmann-DeRidder:2007nzq,GehrmannDeRidder:2007hr,GehrmannDeRidder:2008ug,
      GehrmannDeRidder:2009dp,%
      Weinzierl:2006ij,Weinzierl:2006yt,Weinzierl:2008iv,Weinzierl:2009nz,%
      Currie:2013dwa,Currie:2016ytq,Currie:2017tpe,Currie:2017eqf,
      Britzger:2018zvv,Currie:2018oxh,%
      Bernreuther:2011jt,Bernreuther:2013uma,Chen:2016zbz,Bernreuther:2016ccf,
      Bernreuther:2018ynm,%
      Abelof:2011jv,Abelof:2011ap,Abelof:2012rv,Abelof:2012he,Abelof:2014fza,
      Abelof:2014jna,Abelof:2015lna,%
      Chen:2014gva,Chen:2016zka,Chen:2018pzu,Cruz-Martinez:2018rod,
      Cruz-Martinez:2018dvl,Gehrmann-DeRidder:2017mvr,Niehues:2018was,
      Gehrmann-DeRidder:2018kng},
the CoLoRfulNNLO scheme
\cite{Somogyi:2005xz,Somogyi:2006da,Somogyi:2006db,Aglietti:2008fe,
      DelDuca:2013kw,Kardos:2017xwz,Kardos:2018kqd,%
      DelDuca:2015zqa,DelDuca:2016csb,DelDuca:2016ily,Kardos:2018kqj},
$q_T$-slicing
\cite{Catani:2007vq,%
      Catani:2009sm,Ferrera:2011bk,Catani:2011qz,Gehrmann:2014fva,
      Cascioli:2014yka,Grazzini:2013bna,Grazzini:2015nwa,Bonciani:2015sha,%
      Grazzini:2015hta,Grazzini:2016ctr,deFlorian:2016uhr,Grazzini:2017ckn,
      Grazzini:2017mhc,Grazzini:2018bsd},
$N$-jettiness slicing
\cite{Gaunt:2015pea,%
      Boughezal:2015aha,Boughezal:2015ded,Boughezal:2016wmq,Campbell:2016jau,
      Campbell:2016yrh,Campbell:2016lzl,Moult:2016fqy,Moult:2017jsg,
      Ebert:2018lzn},
sector-improved residue subtraction
\cite{Czakon:2010td,Czakon:2011ve,Czakon:2014oma,%
      Baernreuther:2012ws,Czakon:2012zr,Czakon:2012pz,Czakon:2013goa,
      Czakon:2014xsa,Czakon:2015owf,Czakon:2016ckf,Czakon:2017dip,%
      Boughezal:2011jf,Boughezal:2013uia,Boughezal:2015dra,Caola:2017dug,
      Caola:2017xuq,Rontsch:2018edd,%
      Brucherseifer:2013iv,Brucherseifer:2013cu,Brucherseifer:2014ama,%
      Caola:2014daa},
the Projection-to-Born method
\cite{Cacciari:2015jma,Currie:2018fgr},
Local Analytic Sector Subtraction
\cite{Magnea:2018vyy,Magnea:2018jsj,Magnea:2018hab}
and Geometric IR subtraction
\cite{Herzog:2018ily}.

Here, we discuss two new aspects of the sector-improved residue subtraction
scheme. After reviewing the basic structure of the scheme in
Sec.~\ref{sec:basics}, we present in Sec.~\ref{sec:phasespace} a new phase space
construction with the goal of minimizing the number of distinct subtraction
kinematics. The new phase space construction necessitates the rederivation of
the corrections for the 't~Hooft-Veltman scheme which allows for a
four-dimensional treatment of the resolved particles. We sketch their derivation
in Sec.~\ref{sec:tHV} before we conclude in Sec.~\ref{sec:conclusions}.

\section{Sector-improved residue subtraction scheme}\label{sec:basics}
In order to establish the notation, we briefly review the basic structure of
the sector-improved residue subtraction scheme. The starting point is the
hadronic cross-section for two incoming hadrons $h_{1,2}$ with momenta
$P_{1,2}$, respectively, which can be written as
\begin{align}
  \sigma_{h_1 h_2}(P_1,P_2) &=
    \sum_{a,b} \iint_0^1 \dd{x_1}\dd{x_2} 
    f_{a/h_1}(x_1,\mu_F^2) f_{b/h_2}(x_2,\mu_F^2)
    \hat{\sigma}_{ab}(x_1P_1,x_2P_2;\alpha_s(\mu_R^2),\mu_R^2,\mu_F^2)
  \,,
\end{align}
where $f_{a/h}$ are the parton distribution functions for parton $a$ inside
the hadron $h$. The partonic cross-section $\hat{\sigma}_{ab}$ can be
expanded perturbatively in the strong coupling $\alpha_s$
\begin{align}
  \hat{\sigma}_{ab} &=
    \hat{\sigma}^{(0)}_{ab}+\hat{\sigma}^{(1)}_{ab}+\hat{\sigma}^{(2)}_{ab}
    +\order{\alpha_s^3}
  \,.
\end{align}
At leading order (LO) there is only the Born contribution
$\hat{\sigma}^{(0)}_{ab} = \Sig{B}$, whereas we can distinguish three
contributions at NLO based on the number of final state particles, initial state
convolutions and loops,
\begin{align}
  \hat{\sigma}^{(1)}_{ab} &= \Sig{R} + \Sig{V} + \Sig{C}
  \,.
\end{align}
For the following discussion, we mostly concentrate on the NNLO part
$\hat{\sigma}^{(2)}_{ab}$, for which we find five contributions
\begin{align}
  \hat{\sigma}^{(2)}_{ab} &= \Sig{RR}+\Sig{RV}+\Sig{VV}+\Sig{C1}+\Sig{C2}
  \,.
\end{align}
Schematically, we can write these contributions as
\begin{align*}
  \Sig{RR} &= \frac{1}{2\hat{s}} \int\dd{\Phi}_{n+2}
    \braket{\mathcal{M}^{(0)}_{n+2}}{\mathcal{M}^{(0)}_{n+2}}
	\mathrm{F}_{n+2}
  \,, &
  \Sig{C1} &= \left(\text{single convolution}\right)
          \mathrm{F}_{n+1}
  \,, \\
  \Sig{RV} &= \frac{1}{2\hat{s}} \int\dd{\Phi}_{n+1}
    2\text{Re}\braket{\mathcal{M}^{(0)}_{n+1}}{\mathcal{M}^{(1)}_{n+1}}
    \mathrm{F}_{n+1}
  \,, &
  \Sig{C2} &= \left(\text{double convolution}\right)
          \mathrm{F}_{n}
  \,, \\
  \Sig{VV} &= \frac{1}{2\hat{s}} \int\dd{\Phi}_{n}
    \left(
      2\text{Re}\braket{\mathcal{M}^{(0)}_{n}}{\mathcal{M}^{(2)}_{n}} +
      \braket{\mathcal{M}^{(1)}_{n}}{\mathcal{M}^{(1)}_{n}}
    \right) \mathrm{F}_{n}
  \,,
\end{align*}
where $\dd{\Phi}_{n}$ denotes the $n$-particle phase space integral, $\hat{s}$
is the partonic centre-of-mass energy and $\ket{\mathcal{M}_n^{(\ell)}}$ is the
$\ell$-loop amplitude for the $n$-particle process. The C1 and C2 contributions
contain convolutions with the splitting functions from initial state
factorisation and we refrain from presenting them explicitly for brevity. Their
explicit structure can be found, e.g., in \cite{Czakon:2014oma}.
Finally, each contribution contains a measurement function $F_{n}$, which
defines the observable under consideration (e.g. via jet algorithms, cuts or
histogramming). It has to ensure infrared safety of the observable, i.e. if a
particle of a $n+2$ particle configuration becomes unresolved, $F_{n+2}$ has to
approach $F_{n+1}$ in this limit and similarly for $F_{n+1}$ and $F_n$, which
have to coincide if a particle of the $n+1$ particle configuration becomes
unresolved. Moreover, the measurement functions turn out to be a powerful tool
for the construction of the subtraction scheme in four dimensions, as will be
discussed Sec.~\ref{sec:tHV}.

Since the aim of a subtraction scheme is to make the cancellation of the
infrared singularities explicit, we first have to expose the singularities in
an easily accessible way. The central idea of the scheme is to use sector
decomposition \cite{Binoth:2000ps,Anastasiou:2003gr,Binoth:2004jv} to extract
the divergences. Thus, we have to partition the phase space into sectors and
parametrise each sector such that the divergences occur at the boundaries of
the parameters.
As a first step, to simplify the singularity structure, we split up the phase
space into sectors using a partition of unity, similar to the FKS construction.
For example, the double real phase space can be partitioned using
\begin{align}
  1 &= \sum_{i,j} \left[\sum_k S_{ij,k}
	               +\sum_{k,l} S_{i,k;j,l}\right]
  \,.
\end{align}
Here, each $S_{ij,k}$ singles out the soft singularities of partons
$i$ and $j$ as well as the collinear singularities of any combination of partons
$i$, $j$ and $k$, while it regulates all other infrared singular limits by
tending to zero fast enough. The terms with $S_{ij,k}$ are called
triple collinear sectors. Similarly, $S_{i,k;j,l}$ allows the soft
limits of partons $i$ and $j$ and the pairwise collinear limits of partons $i$
and $k$ and partons $j$ and $l$. These are the double collinear sectors.
In each of these sectors up to two partons ($i$, $j$) are allowed to become
unresolved and we denote their momenta by $u_i$ in the following. One or two
partons ($k$, $l$) take the role of a reference momentum for the collinear
limits to which the unresolved partons can become collinear. We denote their
momenta by $r_i$.

Next, we prepare the factorisation of the double soft limits by ordering the
energies of the unresolved partons in each sector with another partition of
unity,
\begin{align}
  1 &= \theta(u_1^0 - u_2^0) + \theta(u_2^0 - u_1^0)
  \,.
\end{align}
As the infrared singularities correspond to divergences of the matrix elements
when energies or angles between parton momenta vanish, it makes sense to
parametrise the phase space of the unresolved momenta $u_i$ in terms of energy and
angle variables. We choose
\begin{align}
  \hat{\eta}_i &= \frac{1}{2} (1 - \cos\theta_{ir})
  \,, &
        \hat{\xi}_i &= \frac{u_i^0}{u_{i,\text{max}}^0}
  \,,
\end{align}
where $\cos\theta_{ir}$ is the angle between parton $i$ and its reference parton
and $u_{i,\text{max}}^0$ is the maximal energy allowed for parton $i$.
Depending on the sector it may be necessary to introduce a final partition of
unity to be able to remap the collinear divergences to parameter boundaries. The
triple collinear sectors, for example, are decomposed into three subsectors
according to
\begin{align}
  1 &= \underbrace{\theta\left(0 \leq \hat{\eta}_2 \leq \frac{\hat{\eta}_1}{2}
         \right)}_{S_1}
      +\underbrace{\theta\left(0 \leq \hat{\eta}_1 \leq \frac{\hat{\eta}_2}{2}
         \right)}_{S_{23}}
      +\underbrace{\theta\left(\frac{\hat{\eta}_1}{2} \leq \hat{\eta}_2
         \leq \hat{\eta}_1\right)}_{S_4}
      +\underbrace{\theta\left(\frac{\hat{\eta}_2}{2} \leq \hat{\eta}_1
         \leq \hat{\eta}_2\right)}_{S_5}
  \,.
\end{align}
Compared to the construction of \cite{Czakon:2014oma} there is one less
subsector: We have merged subsectors $S_2$ and $S_3$ into one subsector $S_{23}$
like it was suggested in \cite{Caola:2017dug} since the soft and collinear
limits factorise independently.
For each of these subsectors there then exists a reparametrisation of the energy
and angle variables which maps the singularities to the parameter boundary at
zero and factorises overlapping singularities.
As an example, we take the first subsector $S_1$ for which the final
parametrisation reads
\begin{align}
  \hat{\eta}_1 &= \eta_1 \,, &
  \hat{\eta}_2 &= \frac{\eta_1 \eta_2}{2} \,, &
  \hat{\xi}_1  &= \xi_1 \,, &
  \hat{\xi}_2  &= \xi_1 \xi_2 \xi_{2,\text{max}} \,.
\end{align}

At this point it is possible to construct subtraction terms for each sector and
subsector separately by using the soft and collinear factorisation formulae of
QCD, parametrising them with the variables discussed above and applying the
identity (in the distributional sense)
\begin{align}
  \frac{1}{x^{1+b\varepsilon}}
    &= -\frac{\delta(x)}{b\varepsilon}
       +\left[\frac{1}{x^{1+b\varepsilon}}\right]_+
  \,.
  \label{eq:dist-id}
\end{align}
Here, $[f(x)]_+$ denotes plus distributions which are defined as
\begin{align}
  \int_0^1 \dd{x} [f(x)]_+ g(x)
    &= \int_0^1 \dd{x} (f(x)g(x) - f(x)g(0))
  \,.
  \label{eq:plus-func}
\end{align}
This suffices since QCD matrix elements only diverge as
$x_i^{-1-b_i \varepsilon}$ in the energy and angle variables $x_i$. A slight
generalisation of the above prescription is necessary for one-loop matrix
elements since they have several different scaling behaviours at the same time,
see \cite{Czakon:2014oma} for details.
This prescription yields three terms for each singular variable: We call the
first term on the right-hand side of Eq.~\eqref{eq:plus-func} \emph{regular
term}, the second term \emph{subtraction term} and the first term in
Eq.~\eqref{eq:dist-id} \emph{pole term}. Only the pole term contains explicit
poles in $\varepsilon$ and the combination of the regular and subtraction terms
is integrable in the variable $x$.
To obtain the residues of the collinear and soft singularities for the pole and
subtraction terms, we have to employ the standard factorisation formulae of QCD
which are known to at least NNLO in the literature \cite{Aybat:2006mz,
Becher:2009kw,Czakon:2009zw,Mitov:2009sv,Ferroglia:2009ii,Mitov:2010xw,
Becher:2009cu,Catani:1999ss,Campbell:1997hg,Catani:1998nv,Czakon:2011ve,
Bern:1994zx,Bern:1998sc,Kosower:1999xi,Kosower:1999rx,Somogyi:2006db,
Catani:2000pi,Bierenbaum:2011gg,Czakon:2018iev}.

Finally, let us establish some notation in order to further subdivide the
cross-section contributions. A more precise and detailed discussion can be found
in \cite{Czakon:2014oma}.
At NLO, we can split the real contribution $\Sig{R}$ according to regular and
subtraction terms, $\sig{R}{F}$, and pole terms $\sig{R}{U}$ and the virtual
contribution $\Sig{V}$ into a finite remainder $\sig{V}{F}$ and pole terms
$\sig{V}{U}$. Only the finite remainder involves loop matrix elements.
Similarly, we subdivide the NNLO contributions as follows:
\begin{align}
  \Sig{RR} &= \sig{RR}{F} + \sig{RR}{SU} + \sig{RR}{DU}
  \,, &
  \Sig{RV} &= \sig{RV}{F} + \sig{RV}{FR} + \sig{RV}{SU} + \sig{RV}{FR}
  \,, &
  \Sig{C1} &= \sig{C1}{SU} + \sig{C1}{DU}
  \,, \\
  \Sig{VV} &= \sig{VV}{F} + \sig{VV}{FR} + \sig{VV}{F}
  \,, &
  \Sig{C2} &= \sig{C2}{FR} + \sig{C2}{DU}
  \,,
\end{align}
The double real part, $\sig{RR}{F}$, contains the regular and subtraction
terms of the full $n+2$ particle configurations and the single and double
unresolved parts, $\sig{RR}{SU}$ and $\sig{RR}{DU}$, contain pole terms (and
their subtraction terms) which have $n+1$ and $n$ particles configurations,
respectively.
In the case of the real-virtual contribution, we distinguish the full finite
remainder $\sig{RV}{F}$ which contains regular and subtraction terms for
the $n+1$ configuration with one-loop matrix elements, the one-loop finite
remainder pole terms $\sig{RV}{FR}$ which have $n$ particle kinematics but
contain loop corrections, and $n+1$ and $n$ particle tree-level terms
$\sig{RV}{SU}$ and $\sig{RV}{DU}$.
A similar decomposition applies for the double virtual part, where all terms
have $n$ particle kinematics, but $\sig{VV}{F}$ is the full two-loop finite
remainder, $\sig{VV}{FR}$ are the pole terms with one-loop matrix elements and
$\sig{VV}{DU}$ are the pole terms with tree-level matrix elements.
The convolution contribution $\Sig{C1}$ is organised into single and double
unresolved contributions according to the final state multiplicity and for
$\Sig{C2}$ we distinguish parts with one-loop matrix elements ($\sig{C2}{FR}$)
and parts with only tree-level matrix elements ($\sig{C2}{DU}$).

\section{Improved phase space construction}\label{sec:phasespace}
The original phase space construction presented in \cite{Czakon:2010td} was
formulated in conventional dimensional regularisation (CDR) in $d$ space-time
dimensions \cite{tHooft:1972fi,Ashmore:1972uj,Cicuta:1972jf,Bollini:1972ui,
Marciano:1974tv}. It was reformulated in four dimensions using the
't~Hooft-Veltman scheme in \cite{Czakon:2014oma}. This construction still leaves
room for improvements. In particular, the number of distinct kinematic
configurations for the subtraction terms is not minimal. This leads to problems
with misbinning: As in any subtraction scheme, the event and its subtraction
events have different kinematics and, therefore, can contribute to different
bins of histogrammed distributions. This is called misbinning. The
configurations only have to coincide in the singular limits in order to cancel
each other and guarantee integrability. Far away from singular limits misbinning
is of no concern. Close to the singular limits, on the other hand, the weights
of the event and its subtraction events become large and have opposite signs.
Thus, if misbinning occurs close to a singular limit, it can spoil the numerical
convergence of the Monte Carlo integration. Therefore, it is desirable to
minimize the number of distinct subtraction kinematics as this reduces the
probability of configurations ending up in different bins.

With this idea in mind, let us reexamine the basic steps of the phase space
construction from \cite{Czakon:2014oma}. There, the construction of a $n+2$
particle configuration starts with the two unresolved momenta $u_{1,2}$ subject
only to the constraints imposed by the available energy. Afterwards, the
remaining phase space is filled with an $n$ particle configuration for the
corresponding Born process. As an example, let us consider the configurations of
the first triple collinear subsector $S_1$. For each limit, we list the
observable momenta (i.e., soft momenta are removed and collinear momenta have to
be summed) and ellipses denote the remaining momenta of the Born process.
\begin{center}
  \begin{tabular}{ccccc}
    regular &
    $u_2$ soft &
    $u_1,u_2$ soft &
    $r \parallel u_2$ &
    $r \parallel u_1 \parallel u_2$
    \\
    $\{r,u_1,u_2,\dots\}$ &
    $\{r,u_1,\dots\}$ &
    $\{r,\dots\}$ &
    $\{r+u_2,u_1,\dots\}$ &
    $\{r+u_1+u_2,\dots\}$
  \end{tabular}
\end{center}
Since the Born configuration explicitly depends on $u_{1,2}$, these five
configurations do not agree in general.

The idea of the improved phase space construction is now to guarantee a relation
between these limiting configurations. In particular, we want to achieve that
the single unresolved configurations ($\{r,u_1,\dots\}$ and
$\{r+u_2,u_1,\dots\}$) and the double unresolved configurations ($\{r,\dots\}$
and $\{r+u_1+u_2,\dots\}$) agree, thereby reducing the number of distinct
configurations from five to three. To this end, in the new phase space
construction, we start by generating a Born configuration, then add the
unresolved momenta $u_i$ and finally adjust the Born configuration to restore
momentum conservation.

The details of this construction will be presented elsewhere, but the main steps
(for final state references) are summarised here. Similar ideas can already be
found in \cite{Frixione:2002ik,Frixione:2007vw}. In the following $P$ and
$\tilde{P}$ denote the total initial state momentum of the full and the Born
configurations.
We start by postulating a mapping from the $n+2$ particle configuration
$\{P,r_j,u_k,\dots\}$ to the Born configuration
$\{\tilde{P},\tilde{r}_j,\dots\}$ and then specify additional requirements on
the mapping to make it unique:
\begin{itemize}
  \item The mapping must keep the direction of the reference momentum $r$ fixed.
  \item The mapping must be invertible for fixed $u_k$,
        $\{\tilde{P},\tilde{r}_j,u_k,\dots\} \to \{P,r_j,u_k,\dots\}$.
        Thus, once we specify the Born configuration and the $u_k$, we can find
        the $n+2$ particle configuration.
  \item The mapping must preserve the invariant mass of the remaining Born
        configuration, $q^2 = \tilde{q}^2$,
        with $\tilde{q} = \tilde{P} - \sum_{j=1}^{n_{fr}} \tilde{r}_j$
        and $q = P - \sum_{j=1}^{n_{fr}} r_j - \sum_{k=1}^{n_u} u_k$, where
        $n_{fr}$ is the number of reference partons in the final state and $n_u$
        denotes the number of unresolved partons.
\end{itemize}
An algorithm that fulfils these requirements is:
\begin{enumerate}
  \item Generate a Born phase space configuration. One or two of these momenta
        are picked as reference momenta $\tilde{r}_j$ according to the sector we
        consider.
  \item Generate unresolved momenta $u_k$, subject only to constraints on their
        energy. Inserting the unresolved momenta of course violates momentum
        conservation, which has to be restored in the next two steps.
  \item Rescale the reference momenta, e.g., $r = x\tilde{r}$, where the factor
        $x$ can be found from momentum conservation together with the
        requirement $q^2 = \tilde{q}^2$. Rescaling the massless reference
        momentum keeps it on-shell and fulfils the condition of leaving the
        direction unchanged.
  \item Apply a Lorentz transformation to the non-reference final state momenta
        of the Born configuration to restore momentum conservation -- this of
        course preserves their total invariant mass $\tilde{q}^2$.
  \item Multiply the phase space weight by the Jacobian of the transformations.
\end{enumerate}
We basically keep the parametrisations in the subsectors as in
\cite{Czakon:2014oma}, but for subsectors $S_4$ and $S_5$, where the two
unresolved partons $u_1$ and $u_2$ can become collinear to each other, we choose
an energy parametrisation, which parametrises the sum of the energies
$u_1^0+u_2^0$ and their ratio.
Since the relations fixing this procedure (momentum conservation and $q^2 =
\tilde{q}^2$) only depend on the sum (for final state references) or difference
(for initial state references) for the unresolved and resolved momenta, all the
steps above only depend on this sum or difference. In the singular limits we
have $u = \alpha r$. Therefore, the construction keeps $r \pm u$ fixed in these
limits.
For our example, this means that the single soft and single collinear limits
coincide as well as the double soft and the double collinear limit. For the
double unresolved limits the resulting configuration is exactly the underlying
Born configuration with which we started the construction. This is in fact a
general feature: All double unresolved limits correspond to the underlying Born
configuration. This will allow us to discuss the pole cancellation for each Born
phase space point separately in the next section.
Moreover, we expect that this new construction improves the convergence for
invariant mass distributions in absence of final state references, e.g., for
$p p \to t \bar{t}$, since the invariant mass of the top pair is kept constant
across all subtraction configurations.

In order to achieve these features, we have to construct the phase space in the
laboratory frame, while the previous construction was done the CMS frame.
Moreover, the original 't~Hooft-Veltman corrections are spoiled since they depend
on the phase space parametrisation. We discuss their rederivation in the next
section using a new approach.

\section{Rederivation of a four-dimensional formulation}\label{sec:tHV}
In general, it is advantageous to formulate a subtraction scheme in four
space-time dimensions in the sense that the momenta and polarisation vectors of
resolved particles are treated as four-dimensional. One tremendous advantage is
that in this way we can avoid having to use higher orders in $\varepsilon$ of
matrix elements, which would cancel in the final result anyway. The second major
argument for a four-dimensional formulation is the dimensionality of the phase
space integrals. In CDR the number of dimensions that we have to explicitly
parametrise and integrate over grows with the final state multiplicity. In the
't~Hooft-Veltman scheme, where we only treat the unresolved particles as
$d$-dimensional, we never need more than six-dimensional momenta at NNLO.

For many other subtraction schemes, a four-dimensional formulation is relatively
straightforward, since after UV renormalisation dimensional regularisation is
only necessary in order to regulate and cancel the IR singularities. Poles in
the regulator $\varepsilon$ appear explicitly from loop integrals in the virtual
contributions and after phase space integration in the real contributions. If
the unresolved phase space is integrated over analytically, the poles can be
cancelled and the remaining terms can be evaluated in four dimensions. Here,
however, we choose to work with truncated Laurent series in $\varepsilon$ and to
calculate the series coefficients numerically.

When treating the momenta as four-dimensional, we face the challenge that the
only resolved momenta are four-dimensional but the unresolved momenta still
have to be treated in $d$ dimensions. For example, in the single collinear
sector, for $r \parallel u$ the combined momentum $r+u$ must be treated
as four-dimensional, while in the soft limit of $u$ the momentum $u$ is
$d$-dimensional and $r$ is four-dimensional.
If a contribution is free of poles in $\varepsilon$ by construction, like
$\sig{RR}{F}$, $\sig{RV}{F}$ and $\sig{VV}{F}$, we can simply evaluate it
in four dimensions, effectively setting $\varepsilon = 0$. In other
contributions, we can use the measurement functions to force all resolved
momenta to be four-dimensional. We use a replacement along the lines of
\begin{align}
  F_n &\to F_n
    \left(\frac{\mu_R^2 e^{\gamma_\text{E}}}{4\pi}\right)^{-(n-1)\varepsilon}
    \left[\prod_{i=1}^{n-1} \delta^{(-2\varepsilon)}(q_i)\right]
  \,,
  \label{eq:measurement-function}
\end{align}
which forces the $(-2\varepsilon)$-dimensional components of the $q_i$ to zero.
Here, the $q_i$ are the momenta of the resolved final state particles. If the
reference momentum is part of the final state, we have to replace the $q_i$ by
the appropriate resolved combination of momenta, e.g.
$\delta^{(-2\varepsilon)}(r+u)$ in the example discussed above. More details on
this procedure can be found in \cite{Czakon:2014oma}.

By replacing the measurement function as in Eq.~\eqref{eq:measurement-function},
we introduce modifications of order $\order{\varepsilon}$. Therefore, it is
necessary that we only modify combinations of contributions which are separately
finite. This is the case for the combinations $\siG{U} =
\sig{R}{U}+\sig{V}{U}+\Sig{C}$ and $\siG{FR} =
\sig{RV}{FR}+\sig{VV}{FR}+\sig{C2}{FR}$, which also
all come with the $n$ particle measurement function $F_n$. However, for
$\siG{SU} = \sig{RR}{SU}+\sig{RV}{SU}+\sig{C1}{SU}$ and
$\siG{DU} = \sig{RR}{DU}+\sig{RV}{DU}+\sig{C1}{DU}+\sig{VV}{DU}+\sig{C2}{DU}$
the poles only cancel in the combination $\siG{SU}+\siG{DU}$ and $\siG{SU}$
contains terms with both $F_n$ and $F_{n+1}$. This poses a problem to our
procedure since the replacement in Eq.~\eqref{eq:measurement-function} is not
consistent between $F_n$ and $F_{n+1}$. Therefore, it is necessary to first find
correction terms $C_\text{tHV}$ that make $\siG{SU} - C_\text{tHV}$ and
$\siG{DU} + C_\text{tHV}$ separately finite and then apply the replacement to
these combinations separately. Roughly speaking, the idea is to move all
divergent parts of $\siG{SU}$ that come with $F_n$ to $\siG{DU}$ before
applying the 't~Hooft-Veltman scheme.

To find these corrections $C_\text{tHV}$, we start from $\siG{SU}$, which can
be written as
\begin{align}
  \siG{SU}
    &= \sum_{c \in \{RR,RV,C1\}} \int\dd{\Phi_{n+1}} (I^c_{n+1} F_{n+1} + I^c_n F_n)
  \,.
  \label{eq:sigmaSU}
\end{align}
Here, $\dd{\Phi_{n+1}}$ is the phase space integral of the resolved particles.
We use the placeholders $I^c_{n+1}$ and $I^c_n$ for all terms that come with the
$n+1$ and $n$ particle measurement functions, respectively. They can still
contain additional integrations, like convolutions for $C1$ or integrations over
the phase space of the unresolved particle in the case of $RR$.

Let us now consider different choices of measurement functions for a moment. If
we choose a set of measurement functions $\tilde{F}_{\{n,n+1,n+2\}}$ which
requires $n+1$ resolved particles in the final state, we effectively deal with
an NLO calculation for an $n+1$ particle process, which also forces
$\tilde{F}_n \equiv 0$. In this case, all of $\siG{DU}$, $\siG{FR}$ and
$\siG{VV}$ vanish identically since they involve only $n$ particle final
states. The three $SU$ contributions $\sig{RR}{SU}$, $\sig{RV}{SU}$ and
$\sig{C1}{SU}$ take the role of $\sig{R}{U}$, $\sig{V}{U}$ and $\Sig{C}$ of
a $n+1$ particle NLO calculation, respectively. Moreover, only the $I^c_{n+1}$
terms in Eq.~\eqref{eq:sigmaSU} remain and we retain
\begin{align}
  \tilde{\sigma}_\text{SU}
    &= \sum_{c \in \{\text{RR,RV,C1}\}}
       \int\dd{\Phi_{n+1}} I^c_{n+1} \tilde{F}_{n+1}
  \,.
  \label{eq:NLOsigmaSU}
\end{align}
As long as the $\tilde{F}$ measurement functions define an infrared safe
observable, the finiteness of the NLO cross-section implies that the poles in
$\varepsilon$ of $\tilde{\sigma}_\text{SU}$ cancel between the three contributions.
As soon as we change back to the original $F_{\{n,n+1,n+2\}}$ measurement
functions, the poles in $\siG{SU}$ reappear. We learn from this that we only
have to remove the ``non-cancelling'' divergences from $\siG{SU}$ to make it
finite. Additionally, we observe that those poles arise only from terms with
$n$ particle measurement functions. Thus, they can simply be added to
$\siG{DU}$ as discussed above.

The central task in the derivation of the 't~Hooft-Veltman corrections is to
identify and extract the ``non-cancelling'' divergences of $\siG{SU}$. Again,
we make use of the measurement functions for this. We introduce
\emph{parametrised measurement functions},
\begin{align}
  F_{n+1}^\alpha
    &= F_{n+1} \theta\left(\min_{i,j} \eta_{ij}-\alpha\right)
               \theta\left(\min_i \frac{u_i^0}{E_\text{norm}}-\alpha\right)
  \,,
  \label{eq:paramF}
\end{align}
which depend on an auxiliary parameter $\alpha$. The normalisation
$E_\text{norm}$ is arbitrary, but fixed for the whole process. For $\alpha = 0$
the parametrised measurement functions correspond exactly to the original
measurement functions, but for $\alpha > 0$ the step functions cut-off the
double unresolved limits, i.e. whenever the angular parameter $\eta_{ij} =
\frac{1}{2}(1-\cos\theta_{ij})$ between any two massless partons $i$ and $j$
vanishes or if the normalised energy of any massless parton goes to zero. Thus,
for $\alpha > 0$ they correspond to an NLO measurement function.

Since we know that the poles of the $I^c_{n+1}$ terms cancel for NLO measurement
functions, we can subtract a zero from $\siG{SU}$ by subtracting only the
pole terms of $I^c_{n+1}$,
\begin{align}
  \sigma^c_\text{SU} - \mathcal{I}_c^\alpha
    &= \int \dd{\phi_{n+1}} \left(
         I^c_{n+1} F_{n+1} + I^c_n F_n
         -[I^c_{n+1}]_{\varepsilon^{-2},\varepsilon^{-1}} F_{n+1}^\alpha
       \right)
  \,.
\end{align}
This does not change $\siG{SU}$ after summing over the contributions
$c \in \{\text{RR,RV,C1}\}$.
After some rearrangements it is possible to extract the non-cancelling part
\begin{align}
  N^c(\alpha)
    &= \int \dd{\Phi}_{n+1} [I^c_n]_{\varepsilon^{-2},\varepsilon^{-1}} F_n
       \theta\left(\min_{i,j} \eta_{ij}-\alpha\right)
       \theta\left(\min_i \frac{u_i^0}{E_\text{norm}}-\alpha\right)
  \,.
\end{align}
If we take the limit $\alpha \to 0$, logarithmic divergences in $\alpha$ appear.
However, it is possible to analytically extract and cancel them. Afterwards it
is possible to take the limit $\alpha \to 0$ in all remaining terms and thereby
remove the dependence on the auxilliary parameter $\alpha$ entirely. An extended
discussion of this procedure will be given elsewhere. The remaining terms
constitute exactly the sought-after 't~Hooft-Veltman corrections $C_\text{tHV}$.
After subtracting them from $\siG{SU}$ and adding them to $\siG{DU}$,
these contributions are separately finite and can be treated in the
't~Hooft-Veltman scheme.
While this procedure is inspired by slicing methods, it is applied only in the
derivation of the correction terms and no dependence on the parameter $\alpha$
remains in the final result.

\begin{figure}
  \centering
  \raisebox{-0.5\height}{\includegraphics{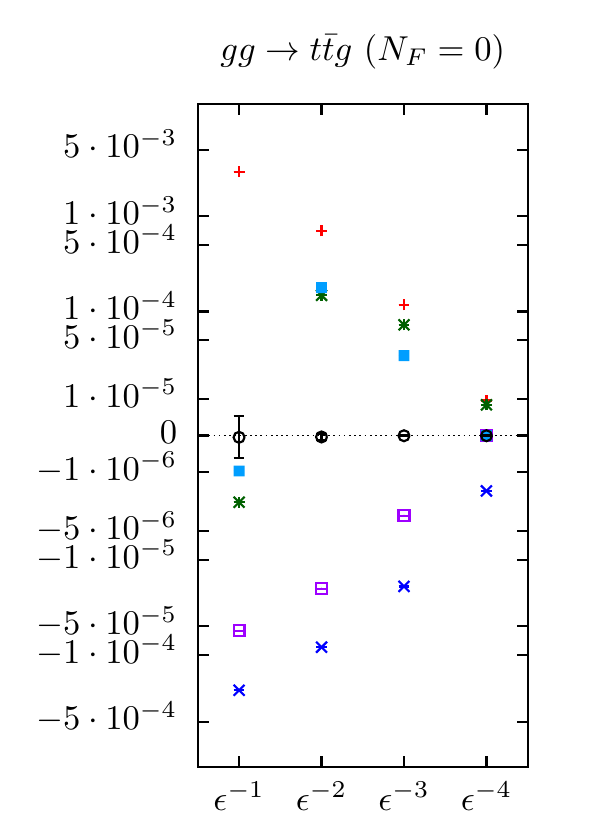}}
  \raisebox{-0.5\height}{\includegraphics{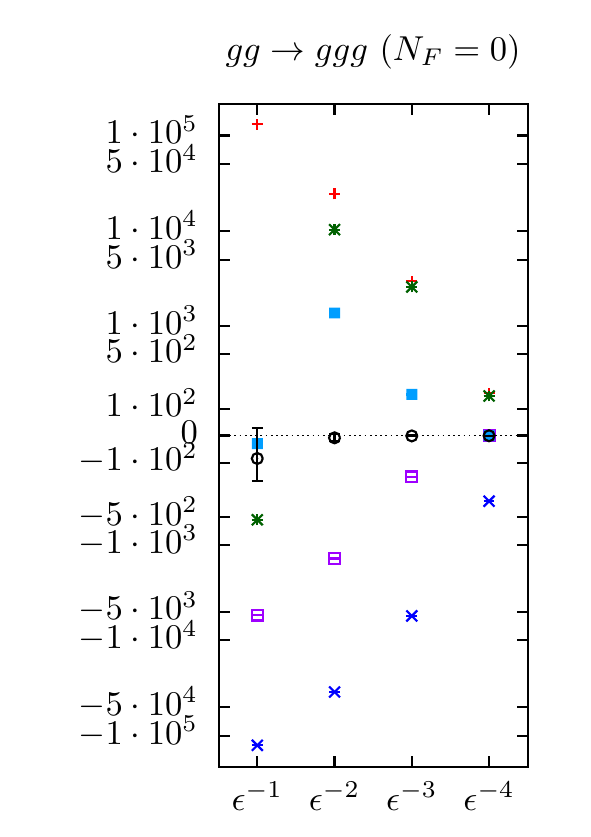}}
  \raisebox{-0.5\height}{\includegraphics{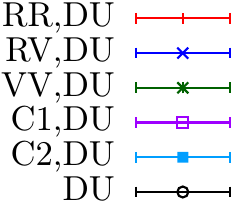}}
  \caption{Illustration of the pole cancellation for the processes
           $gg \to t\bar{t}g$ (left panel) and $gg \to ggg$ (right panel)
           at a fixed phase space point for the underlying Born process.
           For each order in the dimensional regulator $\varepsilon$ the
           size of the individual contributions (see text) is plotted, as
           well as their sum, which is compatible with zero within the
           statistical errors of the Monte Carlo integration over the
           phase space of unresolved particles.}
  \label{fig:pole-cancellation}
\end{figure}
As a demonstration, that this procedure indeed works, let us take the example
processes $g g \to t \bar{t} g$ and $g g \to g g g$ without massless quarks at
NNLO. We  make use of the fact that the new phase space construction allows us
to keep the Born phase space configuration fixed while integrating over the
unresolved phase space of all double unresolved configurations, i.e. up to two
additional unresolved partons.
Fig.~\ref{fig:pole-cancellation} shows the cancellation of the
$\varepsilon^{-4}$ to $\varepsilon^{-1}$ poles for a fixed Born phase space
point. We plot the size of each contribution to $\siG{DU}$ and their sum. The
sum is compatible with zero within the statistical errors of the Monte Carlo
integration, indicated by the error bars. The vertical axis is rescaled using 
$\sinh$ to show both positive and negative contributions at different orders of
magnitude. The values themselves have no physical meaning but only demonstrate
the pole cancellation. This shows that the subtraction scheme presented here
works also for involved massive and massless QCD final states.

\section{Conclusions}\label{sec:conclusions}
In this article we have discussed recent developments in the sector-improved
residue subtraction scheme. We have developed a new phase space construction
which minimizes the number of subtraction kinematics to reduce the probability
of misbinning. Moreover, the scheme allows to easily fix a Born phase space
point while integrating over the unresolved part of the phase space. This can be
used for a pointwise check of pole cancellation. We also rederived the
't~Hooft-Veltman corrections which allow for a four-dimensional treatment of all
resolved momenta and polarisations. Here, we made extensive use of the
measurement functions to derive and calculate the corrections. As an
illustration, we finally considered two involved processes,
$g g \to t \bar{t} g$ and $g g \to g g g$, at NNLO and showed pole cancellation
for a fixed Born configuration.

\bibliographystyle{JHEPM}
\bibliography{lit}

\end{document}